# Observations of Spontaneous Raman Scattering in Silicon Slow-light Photonic Crystal Waveguides

J.F. McMillan[1], Mingbin Yu[2] , Dim-Lee Kwong[2], C.W. Wong[1]
[1]*Optical Nanostructures Laboratory, Columbia University, New York, NY 10027*
[2]*The Institute of Microelectronics, 11 Science Park Road, Singapore, Singapore 117685*

Abstract

We report the observations of spontaneous Raman scattering in silicon photonic crystal waveguides. Continuous-wave measurements of Stokes emission for both wavelength and power dependence is reported in single line-defect waveguides in hexagonal lattice photonic crystal silicon membranes. By utilizing the Bragg gap edge dispersion of the TM-like mode for pump enhancement and the TE-like fundamental mode-onset for Stokes enhancement, the Stokes emission was observed to increase by up to five times in the region of slow group velocity. The results show explicit nonlinear enhancement in a silicon photonic crystal slow-light waveguide device.

The prospect of silicon acting as an active optical material, with the possibility of amplification and lasing, has been the driving force behind the research of Raman scattering in silicon-on-insulator waveguides. This evolved from the first observations of spontaneous Raman scattering in silicon rib waveguides [JalaliOE2002], through the first observations of amplification [JalaliOE2003], and finally to lasing [Jalali2005OE],[Intel2005Nature]. In addition to these achievements in relatively large mode area rib waveguides, the observation of spontaneous scattering [OsgoodOL2003] and amplification [OsgoodOE2004] have been observed in sub-micron channel waveguides. In order to reduce the threshold power of Raman lasers, racetrack cavity [Intel2007OE] lasers have been studied experimentally. It has been shown theoretically that the high confinement and unique dispersion properties of photonic crystal cavities [WongOE2005][WongOE2007] and waveguides [WongOL2006] can be utilized to further reduce threshold values and enhance Stokes emission. Spontaneous Raman scattering has been observed in GaAs photonic crystal slab waveguides [OdaOE2006], however no wavelength dependence of the Stokes emission was reported.

The increase of Raman scattering in photonic crystal waveguides (PhCWGs) when compared to the aforementioned rib waveguide structures is due to two mechanisms: higher modal confinement and larger group indexes. The waveguides studied here have modal area of 0.13 μm$^2$ (averaged over one unit cell in the direction of propagation), which put them close in value to the nanowire waveguide devices studied previously [OsgoodOL2003] [OsgoodOE2004]. Such small modal areas lead to large optical intensities and increase the possibility of Raman scattering. The periodic lattice of the photonic crystal membrane leads to a Bragg-reflection-like lateral confinement (total internal confinement in the vertical direction) for transverse electric (TE) optical mode. This Bragg confinement leads to a uniquely flat dispersion curve at the fundamental mode onset as seen in **Fig. 1a**. In our experiments, we deal exclusively with the fundamental even mode of the waveguide due to the experimental difficulty of effectively exciting the higher-order odd mode. These slow-light frequencies, which have been observed experimentally [NotomiPRL2001], are frequencies where the optical mode experiences large amounts of Bragg reflections from the bulk photonic crystal lattice on the either side of the waveguide. For transverse magnetic (TM) polarized light, no photonic band gap exists [JohnsonPRL1999] however there is still a periodic modulation of the effective index in the direction of propagation due to the lattice [VlasovPRB2006]. This modulation creates a one dimensional photonic crystal, creating a distinctive Bragg stop gap at the Brillouin zone edge. Of course, the corresponding edges of the stop gap exhibit flat dispersion curves, indicative of slow-light. For both of these areas of the dispersion curve, the TE mode onset and the TM stop gap edges, slow-light offers the possibility of increased light-matter interaction, effectively increasing the probability for Raman scattering to take place. It has been shown theoretically that the scattered Stokes intensity in a silicon PhCWGs is inversely proportional to product of the pump and Stokes group velocities [WongOL2006].

In our experiments the pump laser is TM polarized and in the vicinity of the upper edge of the Bragg stop gap. By careful design of the PhCWG parameters (lattice constant, hole radius, slab thickness), the resulting Stokes emission is scattered into the vicinity of the fundamental TE mode onset. This results in *both* the pump and Stokes wavelength having high group indexes and therefore low group velocities.

The waveguides used in our experiment are fabricated utilizing deep-UV photolithography on silicon-on-insulator wafers with a 250 nm thick layer of silicon and a 1 μm layer of $SiO_2$ on silicon substrate. The holes are etched using a plasma dry-etch process. After the dry-etch, a wet etch of hydrofluoric acid is used to remove the underlying $SiO_2$, creating a suspended silicon membrane (see **Fig.1c**). The devices are cleaved manually with no SOI channel waveguides feeding into the PhCWG. This means no control over the exact surface termination of the lattice is achieved in this particular samples [VlasovOL2006]. The fabricated photonic crystal has a lattice constant of 480 nm and holes with a radius of 0.34*a* (163 nm). For this experiment the waveguide length was 1 mm. The fabrication resulted in a hole surface roughness of less than 5 nm and sidewall angle of less than 10 degrees. From cutback measurements the minimum waveguide loss for the TM-like mode was found to be (1.1±0.2) dB/mm and (2.9±1.0) dB/mm for the TE-like mode. In agreement with previous studies [NotomiPRL2005][KraussOE2007] of PhCWG loss, the loss of our waveguides was seen to increase dramatically when approaching the TE mode onset. The transmission properties of the waveguide were experimentally verified using a supercontinuum source and can be seen in **Fig. 1b**.

If any claim of Stokes emission enhancement due to slow group velocity is to be made, the group index of the pertinent modes must be measured. In order to do this, high spectral resolution transmission measurements (wavelength step of 1 pm) were taken of the PhCWG. Utilizing the method outlined in Ref. [NotomiPRL2001], the transmission data of the waveguide was analyzed and the Fabry-Perot oscillation spacing was used to determine the group index. The maximum group index of TM-like and the TE-like mode of the waveguide was measured to be 57 and 149 respectively. The data also shows that there is a 3 nm difference between the desired optical phonon frequency spacing (135 nm, or 15.6 THz) and the measured frequency spacing (138.1 nm) of TM stop band edge with the TE mode onset for the Stokes wavelength. This difference reduces the expected maximum Raman enhancement, which is proportional to product of the Stokes index ($n_s$) and Pump index ($n_p$) [WongOL2006], since the peak $n_s$ doesn't occur at the Stokes wavelength that would be generated by the peak $n_p$ wavelength.

To characterize the spontaneous Raman scattering properties of these waveguides, the experimental setup shown in **Fig.1d** is used. A continuous wave tunable laser source (1530-1565 nm) amplified by an erbium doped fiber amplifier (EDFA) is used as a pump laser. The pump was filtered by a 40 nm wide bandpass filter centered at 1550 nm in order to suppress the EDFA amplified spontaneous emission at the Stokes wavelength. The pump power was monitored with a 99:1 tap, then it was put through a wavelength division multiplexer (WDM) so that the reflected Stokes could be monitored. The pump is coupled into free space via a collimator and then coupled into the waveguide using an aspheric lens (NA = 0.68). The waveguide output was collected using an equivalent lens. A linear polarizer is then used to discriminate between TM and TE polarized light. The light is then collimated back into fiber and the forward scattered Stokes and transmitted pump are separated using another WDM. The Stokes output is then measured using an optical spectrum analyzer (resolution at 500 pm) or a sensitive photodetector. The transmitted pump power is measured using a power meter.

**Fig. 3** shows power dependence of the Stokes emission for a fixed pump wavelength 1535 nm. This wavelength is far from the TM stop band edge, and its corresponding Stokes emission is far from the TE mode onset, therefore the modes of the pump and Stokes emission will be mainly index guided. A simple model of the Stokes emission can be derived [JalaliOE2002][OsgoodOL2004] making it possible to calculate the propagation loss and the Raman scattering efficiency. The model shows that $m_{bak}/m_{fwd} = \sinh(\alpha L)/\alpha L$, where $m_{bak}$ and $m_{fwd}$ are the slopes of the linear fits of the power dependence of the forward and back scattered Stokes emission respectively, $\alpha$ the propagation loss and $L$ the length of the waveguide. From the linear fits, the calculated propagation loss of the waveguide is (2.9$\pm$1.6) dB/mm. In order to calculate the scattering efficiency of the PhCWG, the spontaneous Raman coefficient ($\kappa$) must be calculated using the model which states that $\kappa = m_{fwd} / Le^{-\alpha L}$. From the measured data, $\kappa$ is $(1.9\pm0.9)\times10^{-8}$ $cm^{-1}$, approximately a factor of 3.6 larger than in rib-type silicon waveguides [JalaliOE2002] due to the stronger mode confinement in the PhCWGs. Note also that $\kappa$ is a factor of 2.3 smaller in channel waveguides [OsgoodOL2004], corresponding to the larger modal area in our single line-defect PhCWGs. Knowing $\kappa$ and the solid angle of collection ($\Delta\Omega$ = 0.43 sr), the spontaneous Raman efficiency is calculated to be $(4.2\pm2.0)\times10^{-8}$ $cm^{-1}sr^{-1}$. We emphasize that, in these first observations, $\Delta\Omega$ is not optimized and can be further reduced by controlling the PhCWG surface termination.

To investigate the effect group velocity has on the Stokes emission, the pump wavelength is scanned through the edge of the TM stop band. Due to the fact that the pump wavelength is far from any of silicon's electronic transitions, spontaneous Raman scattering has a very small variation with respect to the pump wavelength. As can be seen in Fig. 4, the Raman emission of the waveguide has a strong pump wavelength dependence which is independent of the wavelength dependant loss of the waveguide. As the pump wavelength approaches the TM stop gap edge, the Raman emission steadily increases, reaching a peak emission which is 5 times higher than that of a wavelength far from the stop gap edge. If the Raman emission variation with pump wavelength was solely dependant on loss one would expect it to drop as it approaches the TM stop gap, as loss in both the TE and TM mode increases as they approach their regions of large group index. The fact that the Raman emission does not decrease as the pump approaches the stop gap edge, but in fact increases is a definitive sign of group index dependence. As can be seen in the data in Fig. 4, the product of the TM and TE mode group indexes follows the Raman emission increase closely. The peak enhancement of the Raman emission is a 5-fold increase (compared to the peaks of the Fabry-Perot). When comparing the increase in the product of the group index, from the start wavelength of 1535 nm to the peak Raman emission wavelength, one can see a 9-fold increase in the product of the group indices.

The difference of the Raman emission enhancement at the slow-light edge from the expected group index product enhancement could be attributed to the group index dependant loss, both linear and nonlinear. It has been shown the nonlinear absorption is group velocity dependent in PhCWGs [BabaCLEO2008]. Though no nonlinear absorption was observed in our power dependence measurements at 1535 nm (away from the slow-light edge), as can be seen by the highly linear power dependence in Fig. 3, it is possible that as the pump wavelength is tuned closer to the slow-light edge, the Raman emission experiences higher losses, thus reducing its ability to reach the maximum expected enhancement.

In conclusion, we have made the first observations of spontaneous Raman scattering in silicon slow-light photonic crystal waveguides. A spontaneous Raman coefficient and efficiency of $(1.9\pm0.9)\times10^{-8}$ $cm^{-1}$ and $(4.2\pm2.0)\times10^{-8}$ $cm^{-1}sr^{-}$

$^{1}$ respectively was measured. In addition, we showed an explicitly group index dependant 5-fold increase in spontaneous Raman scattering. The ability to increase the scattering via slow group velocity structures is encouraging and could be used to create ultra-compact Raman amplifiers and lasers.

Fig.1 – (a) Calculated projected band structure for the W1 photonic crystal waveguide (b) Measured transmission of fabricated waveguide (c) Scanning electron microscope image of photonic crystal waveguide (Scale bar: 1μm) (d) Experimental setup. TL: tunable laser before EDFA and bandpass filter. PC: polarization controller. OSA: optical spectrum analyzer.

Fig.2 – High resolution ($\Delta\lambda$ = 1 pm) tunable laser transmission measurements (solid lines) and derived group index (open circles). (Top) TM-polarization (Bottom) TE-polarization.

Fig.3 – Measured spontaneous Raman Stokes intensity ($\lambda_{pump}$ = 1535 nm). (open squares) Forward scattered (open circles) Backscattered. (Inset) Measured Stokes spectrum for different pump powers.

Fig.4 – (Solid Line) Measured Stokes intensity as a function of pump wavelength (open circles) Product of derived group indexes of the TM-like and TE-like modes. (Inset) Measured Stokes spectrum for $\lambda_{pump}$ = 1535 nm and $\lambda_{pump}$ = 1544.24 nm.

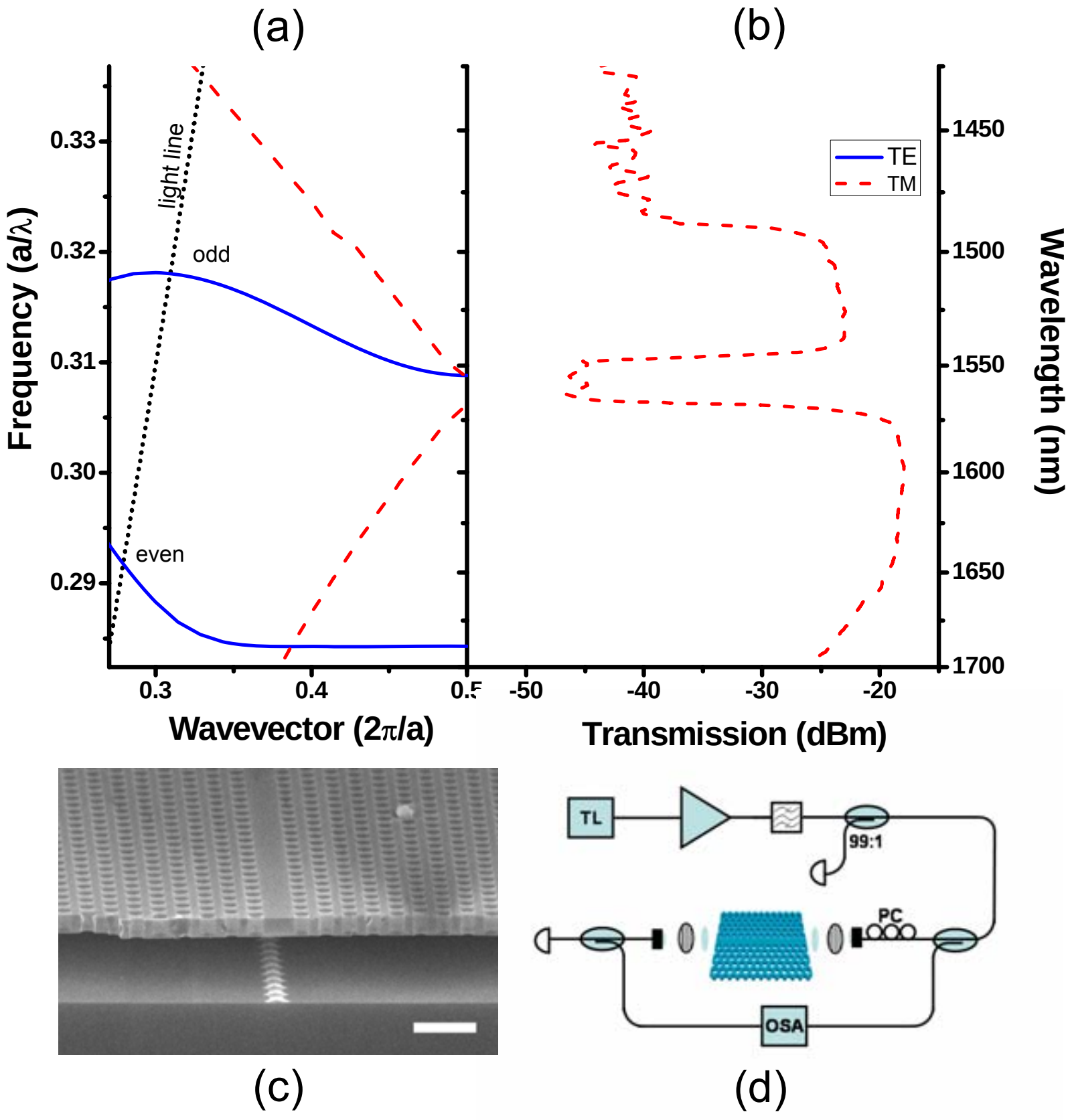

Fig.1 – (a) Calculated projected band structure for the W1 photonic crystal waveguide (b) Measured transmission of fabricated waveguide (c) Scanning electron microscope image of photonic crystal waveguide (Scale bar: 1μm) (d) Experimental setup. TL: tunable laser before EDFA and bandpass filter. PC: polarization controller. OSA: optical spectrum analyzer.

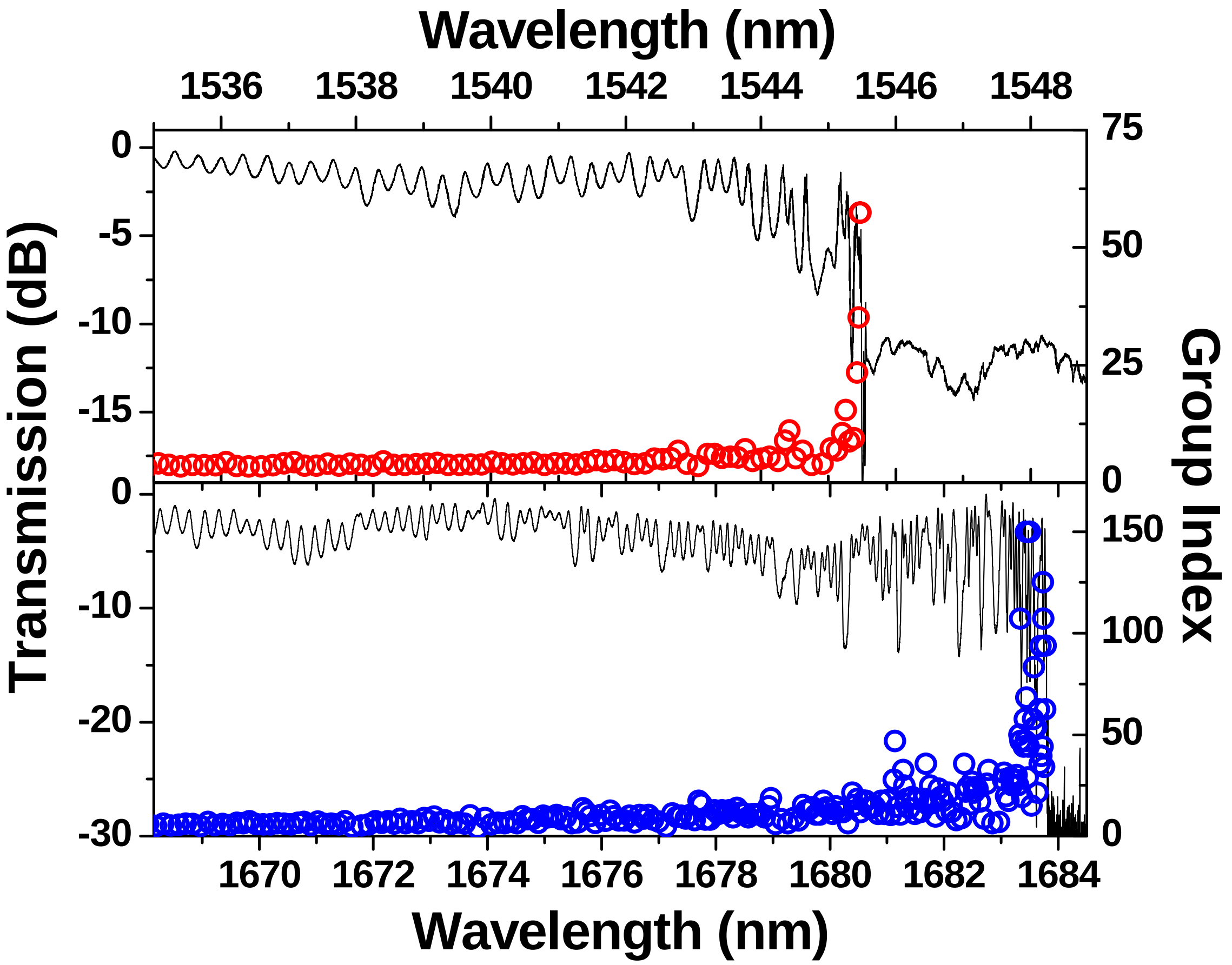

Fig.2 – High resolution ($\Delta\lambda$ = 1 pm) tunable laser transmission measurements (solid lines) and derived group index (open circles). (Top) TM-polarization (Bottom) TE-polarization.

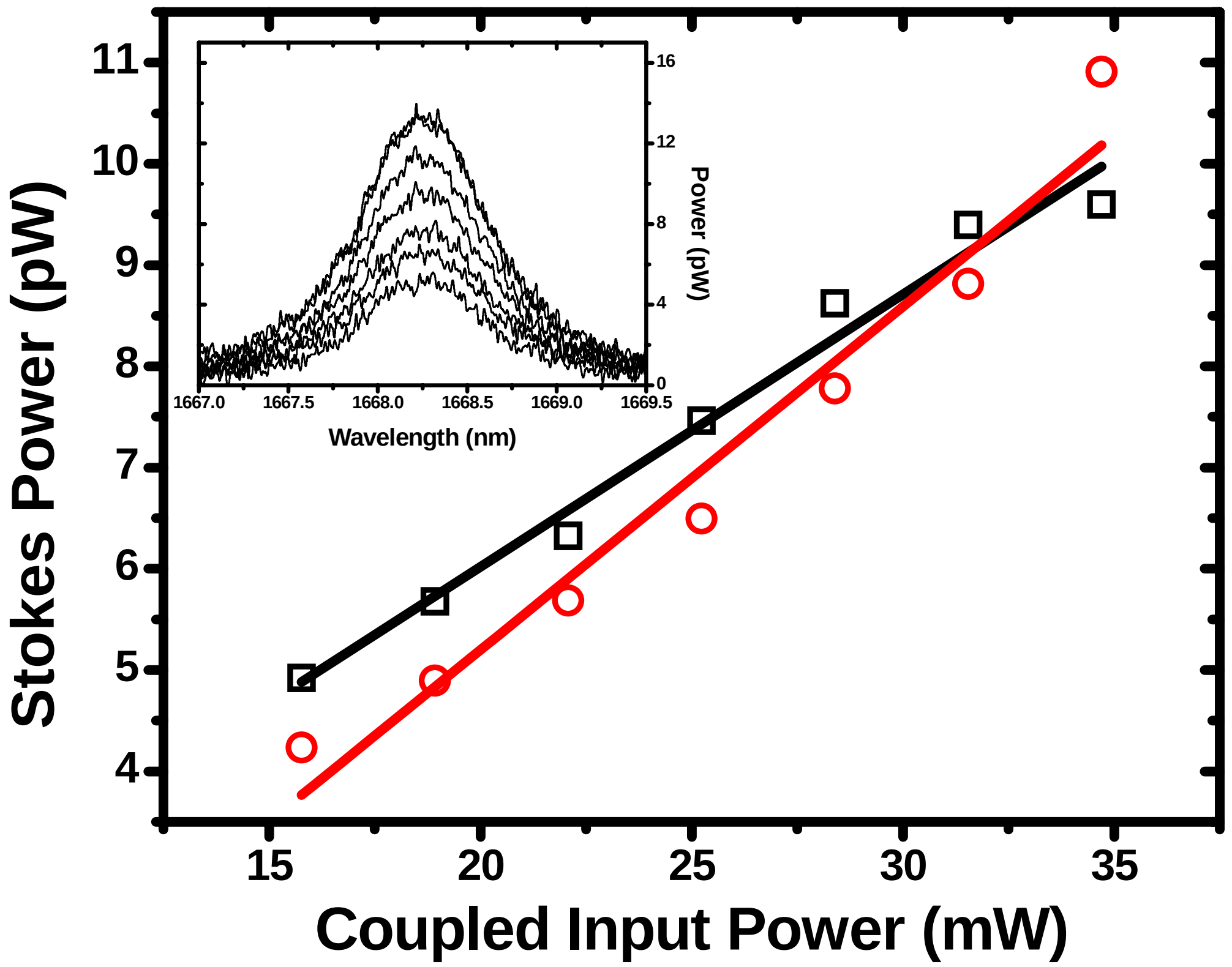

Fig.3 – Measured spontaneous Raman Stokes intensity ($\lambda_{pump}$ = 1535 nm). (open squares) Forward scattered (open circles) Backscattered. (Inset) Measured Stokes spectrum for different pump powers.

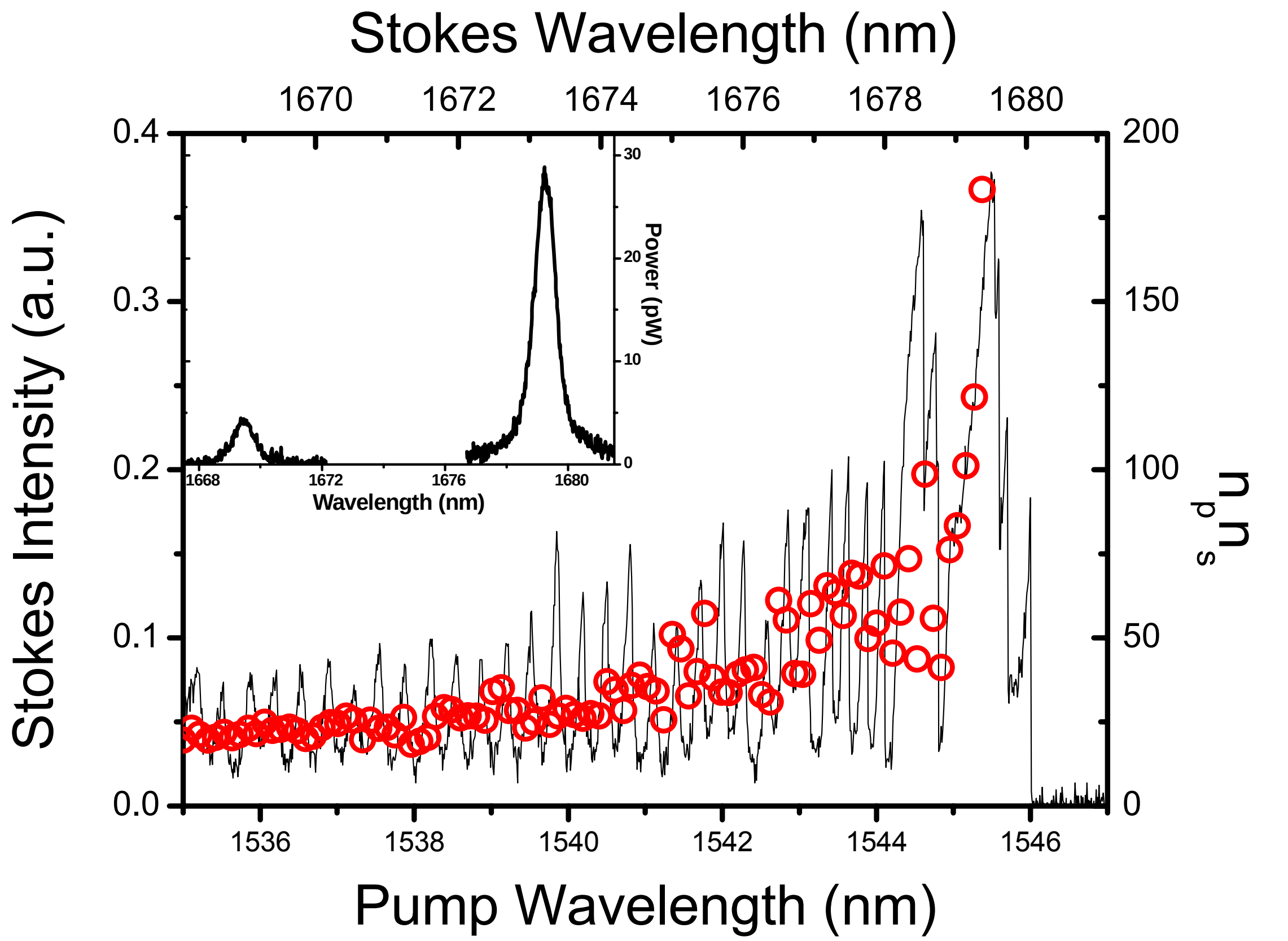

Fig.4 – (Solid Line) Measured Stokes intensity as a function of pump wavelength (open circles) Product of derived group indexes of the TM-like and TE-like modes. (Inset) Measured Stokes spectrum for $\lambda_{pump}$ = 1535 nm and $\lambda_{pump}$ = 1544.24 nm.